\documentclass[a4paper,11pt]{article}

  \usepackage{graphicx}
  \usepackage{amsmath}
  \usepackage{amssymb}
  \usepackage{stmaryrd} 
  \usepackage{MnSymbol}  
  \usepackage[utf8]{inputenc}
  \usepackage[english]{babel}

\begin{document}
 
  \input{DfSliss2e.def}


   \def\dm{\bmth{dm}}          
   \def\rn{\bmth{rn}}          
   \def\tr{\bmth{tr}}          
   \def\sz{\bmth{n}}           
   \def\btc{\bmth{t}}          
   \def\Pr{\bmth{P}}           
   \def\Bool{\bmth{Bool}}      
   \def\true{\mathit{true}}    
   \def\false{\mathit{false}}  
   \def\maxPS{\mathbf{maxPS}}
   \def\tl{\bmth{tl}}          
   \def\wtr{\bmth{wtr}}        
   \def\tm{\bmth{tm}}          
   \def\TF{\bmth{TF}}          
   \def\Hph{\WDH{\vphb}}
   \def\ES{\emptyset}          

   \newcommand{\mdlr}[1]{$#1$}

   \def\Fn{\bmth{F}}           
   \def\fn{\bmth{fn}}          
   \def\Tr{\bmth{Tr}}          
   \def\Inp{\bmth{Inp}}        
   \def\cp{\bmth{cp}}          
   \def\Fnul{\CF^{nul}}        
   \def\len{\mathrm{len}}      
   \def\Out{\bmth{Out}}        
   \def\BGF{\bmth{\GF}}        
   \def\BGL{\bmth{\GL}}        
   \def\Ieq{_n^{-1}}           
   \def\smb{\otriangleup}      
   \newcommand{\Dbr}[1]{\llbracket{#1}\rrbracket}
   \def\smb{\bmth{smb}}
   \def\val{\bmth{val}} 
   \def\lit{\bmth{lit}} 
   \newcommand{\brk}[1]{\bmth{\lsem}#1\bmth{\rsem}} 
   \newcommand{\Anb}[1]{\bmth{\left\langle\right.}{#1}\bmth{\left.\right\rangle}}
   \def\SD{\bmth{\vartriangle}}                
   \def\pn{\bmth{pn}}                          
   \def\gr{\bmth{gr}}                          
   \def\BW{\bmth{\curvearrowright}}     
   \def\ES{\emptyset}                   
   \def\pd{\bmth{\mathit{\Pi}}}         
   \def\pe{\bmth{\mathit{\pi}}}         
   \def\ff{\bmth{\vph}}                 
   \def\d{\bmth{d}}                     
   \def\vl{\bmth{vl}}                   
   \def\ST{\bmth{\prec}}                       
   \def\STE{\bmth{\preccurlyeq}}               
   \def\SS{\bmth{\sqsubset}}                   
   \def\SSE{\bmth{\sqsubseteq}}                
   \def\HA{\textbf{halt}}
   \def\Fl{\bmth{\Ph}}                  
   \def\Flm{\bmth{\Ph_{min}}}           
   \def\TC{\bmth{t}}
   \newcommand{\MTH}[1]{$#1$}

 \thispagestyle{empty}

 \begin{center}  {\Large\bf  On entropic convergence of algorithms\\[2ex] in terms of domain partitions}

 \vspace*{3ex}

  \textbf{Anatol Slissenko}\footnote{Partially supported by  French ``Agence Nationale de la Recherche'' under the project EQINOCS (ANR-11-BS02-004)  and  by Government of the Russian Federation, Grant 074-U01.}
 
 Laboratory of Algorithmics, Complexity and Logic (LACL)

 University Paris-East Créteil (UPEC), France\\
 and\\
 ITMO, St.-Petersburg, Russia
 
 {\sl E-mail:} slissenko@u-pec.fr
 
 \vspace*{3ex} {\it Version of \today}
 \end{center}

 \vspace*{1cm}
 	 \begin{abstract}
 	The paper describes an approach to measuring convergence of an algorithm to its result in terms of an entropy-like function of partitions of its inputs of a given length. The goal is to look at the algorithmic data processing from the viewpoint of information transformation, with a hope to better understand the work of algorithm, and maybe its complexity. The entropy is a measure of uncertainty, it does not correspond to our intuitive understanding of information. However, it is what we have in this area. In order to realize this approach we  introduce a measure on the inputs of a given length based on the Principle of Maximal Uncertainty: all results should be equiprobable to the algorithm at the beginning. An algorithm is viewed as a set of events, each event is an application of a command. The commands are very basic. To measure the convergence we introduce a measure that is called entropic weight of events of the algorithm. The approach is illustrated by two examples.  
 	  \end{abstract}

 \markboth{\hspace*{.5cm}\protect\scriptsize  A.~Slissenko. On entropic convergence of algorithms in terms of domain partitions}
  {\hspace*{.5cm}\protect\scriptsize A.~Slissenko. On entropic convergence of algorithms in terms of domain partitions }

  \vspace*{4ex}
  
  \section{Introduction}\label{SecIntro}
  Intuitively we understand that an algorithm extracts information from its inputs while processing them. So it seems useful to find quantitative measures of this information extraction. It may permit to deepen our vision of complexity of algorithms and problems, and help to design more efficient procedures to solve practical algorithmic problems. Unfortunately, as it was noticed by philosophers many years ago (e.g., see \cite{Floridi2013}) there is no mathematical theory of information that reflects our intuition, and the creation of such a theory is not for tomorrow. However, mathematics has such a notion as entropy, that is a measure of uncertainty about knowledge modeled by probabilistic distributions. And entropy, as well as metric, can be seen as tools to evaluate progress in information processing by algorithm. 
  
  In this paper I describe one way of introducing a probabilistic measure and an entropy-like function for the evaluation of speed on convergence of an algorithm towards its result. 
    
  We start with examples in Section~\ref{SecExamplesOfAlg}. Algorithms are supposed to be defined in a low-level language. We fix the size of inputs, and consider the work of a given algorithm over this finite set. The computations are represented as traces consisting of events. Each event is either an assignment (that we call update, that is shorter) or a guard (the formula in conditional branching). To each event we relate a partition of inputs. These partitions constitute a space to deal with. All this is illustrated by the examples.

  Then in Section~\ref{SecTracesEvents} we describe more formally traces of algorithm and input images of events that permits to describe the algorithm execution in terms of logical literals. This notion is also useful to filter out non-informative events. 
    
  In section~\ref{SecInpPartMeasure}  we introduce input partitions defined by events and a probabilistic measure based on the Principle of Maximal Uncertainty. This principle models the following reasoning. Imagine that the algorithm plays agains an adversary, and this adversary wishes to maximize the uncertainty about the result. That means that all outputs should be equiprobable. And this consideration defines a probabilistic measure. We consider a static measure, i.e., a measure that is not changed with advancing of the algorithm towards its result.
  
  After that we introduce an entropic weight of event partitions, and in terms of this weight we evaluate entropic convergence of algorithms from our examples in Section~\ref{SecAnlzExmp}.
  
   In Conclusion we mention strong and weak points of the present approach and what can be done next. 
  
  We use the following notational conventions: an algorithm considered in the general framework is $\GA$, it computes a total function $\Fn$ of bounded computational complexity. For better intuition one may think that problems we consider are not higher than $\bmth{NP}\cup{}\bmth{coNP}$.
   Concrete functions in examples are boldface greek letters; an algorithm computing  function $\Fn$ is denoted $\GA(\Fn)$ or $\GA_m(\Fn)$ if we consider several algorithms that compute $\Fn$. Other notations used in the next section:  $\F_m$ is the finite ring modulo $m$, $\Z$ is the set of integers, $\N$ is the set of natural numbers, $\B=\{0,1\}$. Other notations are introduced in appropriate places. 
  
  We consider only functions $\Fn$ whose output consists of one component, like in the examples below. Functions like convolution, sorting are multi-component, i.e., an algorithm that computes such a function outputs several values written in different locations.
  
  Very brief description of basic constructions of this paper is in \cite{Sli:2016:OECADg}.

 \section{Examples of algorithms}\label{SecExamplesOfAlg}
 
 The following two examples are used to illustrate the approach. 
 We use logical terminology for algorithms\footnote{This terminology has the flavor of the classification of functions introduced by Yu.~Gurevich for his abstract state machines. However our context is quite different from his machines.}, so what are variables in programming are dynamic  functions in our context. We name different objects in our examples as `update', `guard',  `event', `input' etc., though general definitions will be given in the next section~\ref{SecTracesEvents}.
 In particular, the inputs are external functions that may have different values (i.e., they are dynamic) and cannot be changed by the algorithm. But the algorithm can change its internal functions.  Without loss of generality, the output function is supposed to be updated only once to produce the result. The symbol \% introduces comments in algorithm descriptions.
 

\begin{exmp}\label{exmpSumMod2}\rm \textbf{Sum over $\F_2$ or XOR: $\sib$.} First we formulate the problem, and then an algorithm that solves it.

 \VSon{}\underline{Input}: A word $x$ over an alphabet $\B$ of length $n$, i.e., $x\in\B^n$,  we assume   that $n$ even for technical simplicity;  $\nu\DF\frac{n}{2}$.
 
 \VSon{}\underline{Output}: $\sib(x)=\sum_{1\leq{}i\leq{}n}{}x(i)\mod2$. 
 
 Algorithm: a simple loop calculating $\si_j=\sum_{1\leq{}i\leq{}j}{}x(i)$. 
 
  \VSon\NI\underline{Algorithm $\GA(\sib)$}\nopagebreak
 
 \% $x$, $n$ are inputs, $\si$ is output, $i$ is a loop counter, $s$ is an intermediate value 
 
 \% Functions,  $x$, $n$ are external, and the others are internal.\\
 \NI\LNo{1}\Lze $i:=0$; $s:=0$;            \Hc{7.1}\%Initialization\\
 \LNo{2}\Lze \IF\ $i<n$ \TH\ $i:=i+1$; $s:=s+x(i)$; \GO\ 2\\
 \LNo{3}\Lze \Hc{1.4}\EL\   $\si:=s$; \HA\           \Hc{5.1}\% case $i\geq{}n$ \

\VSon{}All traces of $\GA(\sib)$ are 'symbolically' the same (the algorithm is oblivious):

  $i:=0,\;s:=0,\;i<n,\;i:=i+1,\;s:=s+x(i),\;i<n,\;i:=i+1,\;s:=s+x(i),\dots,$ 
   
  $i<n,\;i:=i+1, s:=s+\dots+x(i),\;i\geq{}n,\;\si:=s$
  
  \NI{}Here $i:=0,\;s:=0$ etc. are updates, and $i<n$ and similar are guards that are true in the trace. Thsi is a symbolic trace. Replace internal functions in guards and right-hand side of updates by their values, and we  get a more clear vision of a trace: 
  
  $i:=0,\;s:=0,\;0<n,\;i:=0+1,\;s:=s+x(1),\;1<n,\;i:=1+1,\;s:=s+x(2),\dots,$ 
   
  $n-1<n,\;i:=n-1+1, s:=s+\dots+x(n),\;n\geq{}n,\;\si:=s$  
  
  Let us fix an input, i.e., a value of $x$, and denote the values of its components $[x(i)]$, 
  $1\leq{}i\leq{}n$. Transform the trace for this input into a sequence of literals: replace internal functions by their 'symbolic images' (defined in section~\ref{SecTracesEvents}) in the guards and in the left-hand sides of updates, and replace the right-hand side of updates by their values (this is not formal but self-explanatory):
  
  \NI\Hc{.3}$0=0,\;0=0,\;0<n,\;1=1,\;x(1)=[x(1)],\;1<n,\;2=2,\;x(1)+x(2)=[x(1)]+[x(2)],\dots,n-1<n,$ 
   
  $\;n=n, x(1)+\dots+x(n)=[x(1)]+\dots+[x(n)],\;n\geq{}n,\;x(1)+\dots+x(n)=[x(1)]+\dots+[x(n)]$
   \ESqr
 \end{exmp}

    \begin{exmp}\label{exmpMaxPS}\rm  \textbf{Maximal prefix-suffix ($\maxPS$): $\vphb$}.\\ 
\VSon{}The maxPS problem is simple: given a word over alphabet $\A$, find the length of the maximal (longest) prefix, different from the entire word, that is also a suffix of the word.

 \VSon{}\underline{Input}: A word $w$ over an alphabet $\A$, $\al\DF|\A|\geq2$ of length $n$.
 
 \VSon{}\underline{Output}: $\ff(n,w)=\ff(w)=\max\{k:0\leq{}k\leq(n-1)\wedge{}w(1..k)=w(n-k+1..n)\}$. 
 
  \VSon{}We consider two algorithms for $\maxPS$: a straightforward one $\GA_0(\ff)$ with complexity 
 $\BO{n^2}$, and another one $\GA_1(\ff)$ with complexity $\BO{n}$. The first one is trivial, the second one is  simple and well known. (In the descriptions of algorithms below we  aline $\EL$ with $\IF$, not with $\TH$, in order to economize the space.) 
 
 \VSon\underline{Algorithm $\GA_0(\ff)$}
 
 \NI\LNo{1}\Lze $h:=0$;             \Hc{5.8}\%initialization of the external loop\\
 \LNo{2}\Lze \IF\ $h\geq(n-1)$ \TH\ $\vph:=0$; \HA; 
                                      \Hc{1.5}\%here $\vph$ is a nullary output function\\
 \LNo{3}\Lze \EL  \Hc{6.3}\%  case $h<(n-1)$\\  
 \LNo{4}\Lon \BE\\ 
 \LNo{5}\Ltw $h:=h+1$; $i:=1$; \\ 
 \LNo{6}\Ltw \IF\ $w(i)=w(i+h)$ \TH\\  
 \LNo{7}\Lth  \big(\IF\ $i<n-h$\ \TH\ $i:=i+1$; \GO\ 6;\\
 \LNo{8}\Lfo \EL\   $\vph:=n-h$; \HA;\big)              \Hc{2.6}\% case $i\geq(n-h)$, i.e., $i=(n-h)$\\
 \LNo{9}\Ltw \EL\  \GO\ 2                                \Hc{4.8}\% case $w(i)\neq{}w(i+h)$ \\
 \LNe \Lon \EN
  
  \VSon{}Algorithm $\GA_1(\ff)$ recursively calculates  $\ff(m,w)$ for all $m$ starting from $m=1$. 
  Denote by $\ff^{k}(m)$ the $k$th iteration of $\ff(m)$, $k\geq1$: $\ff^{1}(m)=\ff(m)$ and 
  $\ff^{k+1}(m)=\ff(\ff^{k}(m))$, and assume that $\vph^{0}(m)=-1$ for all $m$, $\vph(0,w)=0$ and 
  $\min\emptyset=0$. 
  
 Denote by letter $\vph$ (not boldface) an internal function of $\GA_1(\ff)$ of type $[0..n]\IM[0..n-1]$, i.e., an array, that represents $\vph(w,m)$ as $\vph(m)$. Its initial value is $\vph(0)=0$. 
 
  \VSon{}Suppose that $\vph(m)$ is defined, and $m<n$. Algorithm $\GA_1(\ff)$ computes 
 
 $\vph(m+1)$ as $\vph^{s}(m)+1$, where 
 $s=\min\{k:w(\vph^{k}(m)+1)=w(m+1)\}$. 
 
 \VSon{}\NI{}Clearly, this computing of $\vph^{s}(m)$ takes 
 $\BO{s}$ steps. The whole complexity of $\GA_1(\ff)$ is linear. 
  
  \pagebreak
\VSfo{}\underline{Algorithm $\GA_1(\ff)$}

\NI\LNo{1}\Lze $i:=1$; $\vph(1):=0;\;\psi:=0$;                         \Hc{4,7}\%initialisation; \\
\LNo{2}\Lze\IF\ $i\geq{}n$ \TH\ \big($\vph(n)=r:=\psi$; \HA\big);  
                                               \Hc{2.1}\% by $r$ we denote our standard output;\\
\LNo{3}\Lze\EL\  \big($i:=i+1$;              \Hc{5.6}\% case $i<n$ \\
\LNo{4}\Ltw\IF\ $w(\psi+1)=w(i)$ 
                         \TH\ \textbf{(} $\vph(i):=\psi+1;\;\psi:=\psi+1$; \GO\ 2\textbf{)}\\
\LNo{5}\Ltw\EL\             \Hc{7}\% case $w(\psi+1)\neq{}w(i)$ \\
\LNo{6}\Lfo\IF\ $\psi>0$ \TH\ $\psi:=\vph(\psi)$; \GO\ 4\\
 \LNo{7}\Lfo\EL\ \GO\ 2 \big)                       \Hc{5}\% case $\psi=0$\\

  \VStw{}Consider the work of algorithms $\GA_0(\ff)$ and $\GA_1(\ff)$ on the input $w_1=a^{n-1}b$ (the traces are given in the next section~\ref{SecTracesEvents}). 
  
  Compare the datum 
$w(n-1)\neq{}w(n)$ obtained by any of these algorithms and the knowledge behind this datum for arbitrary words. One can easily conclude that $w(n-1)\neq{}w(n)$ is possible only for words of the form $w_1$, and this inequality immediately implies that $\vph(w_1,n)=0$. 
However, none of these algorithms outputs the result, they continue to work. The question is what information they are processing, and how they converge to the result. 
  \ESqr

    \end{exmp}

 \section{Traces of algorithms and event partitions}\label{SecTracesEvents}
 
 In our general framework we consider sets of traces, that can be viwed as sets of sequences of commands. One can take traces abstractly, so we do not need too detailed notion of algorithm. However, in order to relate the general setting with the examples more clearly, we make precisions on the representation of algorithms. 
 
  An algorithm $\GA$ is defined as a program over a vocabulary $\V$. 
  
  This vocabulary consists of sets and functions (logical purism demands to distinguish symbols and interpretations but do not do it). The sets are always pre-interpreted, i.e., each has a fixed interpretation: 
 natural numbers $\N$, integers $\Z$, rational numbers $\Q$, elements of finite ring $\F_m$, alphabet 
 $\B=\{0,1\}$, alphabet $\A$, Boolean values $\Bool$, words over one of these alphabets of a fixed length. Elements of these sets are \emph{constants} (from the viewpoint of logic their symbols are nullary static functions). We assume that the values of functions we consider are constants. We also assume that the length of these values is bounded by $\log\sz+\BO{1}$, where $\sz$ is the input length introduced just below. This permits to avoid some pathological situations that are irrelevant to realistic computations, though this constraint are not essential for our examples. 
 
 The functions are classified as \emph{pre-interpreted} or \emph{abstract}. Pre-interpreted functions are: addition and multiplication by constants over $\N$, $\Z$ and $\Q$, operations over $\F_m$, Boolean operations over $\Bool$,    basic operations over words if necessary. Notice that symbols of constants are also pre-interpreted functions. The vocabularies used in our examples are more modest, we take richer vocabularies for further examples that are under analysis. 
  
  Abstract functions are \emph{inputs}, that are external, i.e., cannot be changed by $\GA$, and \emph{internal} ones. We assume that in each run of $\GA$ the output is assigned only once to the output function, and just at the end, before the command $\HA$.   
   Notice that what is called variable in programming is a \emph{nullary} function in our terminology, a 1-dimensional array is a function of arity $1$ etc. The arguments of an internal function serve as index (like, e.g.,  the index of a 1-dimensional array). 
     
  \VSon{}Terms and formulas are defined as usually. 
  
  \VSon{}Inputs, as well as outputs of $\GA$ are sets of substructures over $\V$ without proper internal functions. For inputs and output there is defined \emph{size} that is polynomially related to their bitwise size (e.g., the length of a word, the number of vertices in graph etc.). We fixe the size and denote it $\sz$. For technical simplicity and without loss of generality we consider the inputs of size exactly $\sz$. 
 
  \VSon{}As it was mentioned above, the function computed by $\GA$ is denoted $\Fn$. Its domain, constituted by inputs of size $\sz$,  is denoted 
  $\dm(\Fn)$ or simply $\dm$. The image (the range) of $\Fn$ is denoted $\rn(\Fn)$ or $\rn$; $\rn=\Fn(\dm)$. Variables for inputs are $X,\;Y$ maybe with indices. 
  
  The worst case computational complexity of $\GA$ is denoted $\btc$, and the complexity for a given input $X$ is denoted by $\btc(X)$. We write $t\IM\infty$ instead of $t\IM\btc$ or $t\IM\btc(X)$.
 
 \VSon{} Two basic commands of $\GA$ are guard verification and update; the command $\HA$ is not taken into consideration in traces. A \emph{guard} is a literal (this does not diminish the generality), and an \emph{update} (assignment) is an expression of the form $g(\Th):=\et$, where $g$ is an internal function, $\Th$ is a list of terms matching the arity of $g$, and $\et$ is a term. 
  
  A program of $\GA$ is constructed by sequential composition from updates, branchings of the form 
  $\IF\,guard\,\TH\;Op\;\EL\;Op'$, where $Op$ and $Op'$ are  programs, $\GO\;label$ or $\HA$. 
  
  Given an input $X$, a \emph{trace} of $\GA$ for $X$ denoted $\tr(X)$, is a sequence of updates and guards that correspond to the sequence of commands executed by $\GA$ while processing $X$. More precisely, the updates are the updates executed by $\GA$, and the guards are the guards that are true in the branching commands. So such a guard is either the guard that is written in $\IF$-part or its negation. These elements of a traces are called \emph{events}. The commands $\HA$, $\GO$ and other commands of direct control, are not included in traces, so the last event of a trace is an update of the output function. The event at instant $t$ is denoted by $\tr(X,t)$. 
  
    \VSon{}We assume that the values of internal functions are assigned by $\GA$, and are defined when used in updates. In other words, there are no initial values at instant $0$ (or we can say that all these functions have a special value $\natural$, meaning $undefined$, that is never assigned later), all internal functions are initialized by  $\GA$. This means, in particular, that the first update is necessarily by an `absolute' constant or by an input value. As it was mentioned above,  all values are constants that are external functions.  
  
  Everywhere below $\Th$ in expressions like $f(\Th)$, is a list $\ta_1,\dots,\ta_m$ of terms whose number of elements is the arity of  $f$. 
  
  The value of a term $\th$ in a trace $\tr(X)$ at instant $t$, denoted $\th[X,t]$,  is defined straightforwardly  as follows: 
  
  $\bullet$ if $\ga$ is an external function then its value for any value $\bmth{\Th}$ of its argument $\Th$ is already defined for a given input  $X$, independently of time instant, and is denoted $\ga(\bmth{\Th})[X]$ or 
  $\ga(\bmth{\Th})[X,t]$ to have homogenous notations.
  
  $\bullet$ if $\th=\ga(\Th)$, where $\ga$ is an external function then
   
  \Hc{2}$\th[X,t]=\ga(\Th[X,t])[X]=\ga(\ta_1[X,t],\dots,\ta_m[X,t])[X]$; 
      
  $\bullet$ if $\th=g(\Th)$, where $g$ is an internal function, and 
  if $\th$ is not updated at $t$ then 
  
  \Hc{2}$\th[X,t]=\th[X,t-1]=g(\ta_1[X,t-1],\dots,\ta_m[X,t-1])[X,t-1]$,
  
  \NI{}and if $\tr(X,t)$ is an update $g(\Th):=\et$ then $g(\Th[X,t-1])[X,t]=\et[X,t-1]$ (an update defines $g$ for some concrete arguments that should be evaluated before the update).

   \VSon{}\emph{Input image} of a term $\th$ at $t$ in $\tr(X)$, 
  denoted $\th\Ang{X,t}$, is defined by recursion over time $t$ and term construction: 
  
  $\bullet$ for a term $\ga(\Th)$, where $\ga$ is an external function,  we set 
     $\ga(\Th)\Ang{X,t}=\ga(\Th\Ang{X,t})$ for all $X$ and $t$; 

    $\bullet$ for $g(\Th)$, where $g$ is a internal function and $\tr(X,t)$ is not an update  
  $g(\Th[X,t-1]):=\et$, we set  $g(\Th[X,t-1])\Ang{X,t}=g(\Th[X,t-1])\Ang{X,t-1}$;
  
$\bullet$ for $g(\Th)$, where $g$ is a internal function and $\tr(X,t)$ is  an update  
  $g(\Th[X,t-1]):=\et$, we set  $g(\Th[X,t-1])\Ang{X,t}=\et\Ang{X,t-1}$.
  
  One can see that input image of $g(\Th)$, where $g$ is a internal function, is a term related to $g$ with a concrete argument, i.e., to some kind of nullary function. We can treat the only output in some special way, and we do it later, in order not to loose its trace. 
  
  Logical purism demands that for constants we distinguish the symbol and the value. So for a loop counter $i$ with updates $i:=0$, $i:=i+1$, $i:=i+1$ we get as input images of $i$ the terms $\bmth{0}$, $\bmth{0+1}$ and
  $\bmth{(0+1)+1}$, where boldface refers to symbols. 
  
  \begin{prop}\label{propInpIm1}Input image of a term does not contain internal functions (i.e., is constructed from pre-interpreted functions and inputs). 
  \end{prop}
 \NBF{Proof.} By straightforward induction on the construction of input image. \Qed
  
  \VSon{}\emph{(Trace) literal of an event} $E=\tr(X,t)$ is denoted $E\Ang{X,t}$ or $\tl(X,t)$ (notice that an event may have many occurrences in the traces) and is defined as follows:
  
  $\bullet$ if $E$ is an update $g(\Th):=\et$ and $g$ is not output then $E\Ang{X,t}$ is the literal
   
  \Hc{1}$g(\Th[X,t-1])\Ang{X,t}=\et[X,t]$;  
 
  $\bullet$ if $E$ is an update $g(\Th):=\et$ and $g$ is an output function then as $E\Ang{X,t}$ we take the literal   $g(\Th[X,t-1])=\et[X,t]$;  
 
  $\bullet$ if $E$ is a guard $P(\Th)$  then $E\Ang{X,t}$ is the literal $P(\Th\Ang{X,t})$;
  
  $\bullet$ if $E'=\tr(X,t')$ with $t'<t$  then $E'\Ang{X,t}=E'\Ang{X,t-1}$.
    
 \VSon{}For the example of loop counters $i:=0$, $i:=i+1$, $i:=i+1$ we get as trace literals 
  $\bmth{0}=0$, $\bmth{0+1}=1$ and  $\bmth{(0+1)+1}=2$. These literals are often not instructive for the convergence of $\GA$ to its result. 
  
  Trace literals not containing input functions are \emph{constant trace literals} (parameter $\sz$ is treated as a constant that does not depend on other inputs).
  
  In further constructions, as we illustrate in the examples just below, we do not distinguish symbols and values of  constants, and write, e.g.,  $0+1+1=2$ instead of $\bmth{(0+1)+1}=2$. Moreover, instead of a sum of $1$'s taken, say $m$ times, we write simply $m$ or $(m-1)+1$ according to  the context (that always permits to understand what is meant by this notation).
     
   \VStw{}For algorithm $\GA(\sib)$ of Example~\ref{exmpSumMod2} we have  the following trace literals corresponding to the trace given in this example (we denote the value of an input function $x(i)$ for a concrete $i$ by $[x(i)]$; this value does not depend on $t$ but only on $i[X,t]$):
   
   $0=0,\;0=0,\;0<n,\;0+1=1,\;x(1)=[x(1)],\;1<n,\;0+1+1:=2,\;x(1)+x(2)=[x(1)]+[x(2)],\dots,$ 
   
  $n-1<n,\;n=n,\; x(1)+\dots+x(n)=[x(1)]+\dots+[x(n)],\;n\geq{}n,\;\si=[x(1)]+\dots+[x(n)]$ 
   
   Notice that though we do not replace output event by its 'regular' trace literal (it is done in order to have a reference to the result),  the input image of $\si$ is  $\si\Ang{X,\infty}=x(1)+\dots+x(n)$.

\VStw{}The trace of $\GA_0(\ff)$ of Example~\ref{exmpMaxPS} for input $w_1\DF{}a^{n-1}b$ with $a\neq{}b$ has the form (in order to facilitate the reading we put the current or acquired value $v$ of a term $\th$ behind it 
as $\th$[v]):

 $h:=0,\,h<(n-1),\,h[1]:=h+1,\,i:=1,\,w(1)=w(2),\,i[1]<(n-1),\,i[2]:=i+1,\,w(2)=w(3),\dots,$ 
 
 $w(n-2)=w(n-1), i[n-2]<(n-1),\,i[n-1]:=i+1,\,w(n-1)\neq{}w(n),\,h[1]<(n-1),\,h[2]:=h+1,\dots$
 
 $w(1)\neq{}w(n),\,h[n-1]\geq(n-1),\,\vph:=0$

 \VSon{}The respective trace literals are (denote this sequence $\tl_0(w_1)$):
 
 $0=0,\,0<(n-1),\,1=1,\,1=1,\,w(1)=w(2),\,0+1<(n-1),\,1+1=2,\,w(2)=w(3),\dots,$
 
 $w(n-2)=w(n-1), (n-3)+1<(n-1),\,(n-2)-1=n-1,\,w(n-1)\neq{}w(n),\,1<(n-1),\,2=2,\dots$
 
 $w(1)\neq{}w(n),\,n-1\geq(n-1),\,\vph=0$

 \VStw{}The trace of $\GA_1(\ff)$ of Example~\ref{exmpMaxPS} for input $a^{n-1}b$ with $a\neq{}b$ has the form :
 
 $i:=1,\,\vph(1):=0,\,\ps:=0,\,i<n,\,i:=i+1[2],\,w(1)=w(2),\,\vph(2):=\ps+1[1],\,\ps:=\ps+1[1],$
 
 $i<n,\,i:=i+1[3],w(2)=w(3),\,\vph(3):=\ps+1[2],\,\ps:=\ps+1[2],\dots,i[n-2]<n,$
   
 $i[n-2]:=i+1[n-1],\,w(n-2)=w(n-1),\,\vph(n-1):=\ps+1[n-2],\,\ps:=\ps+1[n-2],$
 
 $i[n-1]<n,\,i:=i+1[n],\,w(n-1)\neq{}w(n),\,\ps[n-2]>0,\ps:=\vph(n-2)[n-3],\,w(n-2)\neq{}w(n),$ 
 
 $\ps[n-3]>0,\,\ps:=\vph(n-3)[n-4],\,w(n-3)\neq{}w(n),\dots,\ps[1]>0,\,\ps:=\vph(1)[0],\,w(1)\neq{}w(n),$
 
 $\ps\leq0,\,i[n]\geq{}n,\, \vph(n):=0,\,r:=0$
 
 \VSon{}The sequence of trace literals of this trace (denote it $\tl_2(w_1)$) is:
 
 $1=1,\,0=0,\,0=0,\,0<n,\,1+1=2,\,w(1)=w(2),\,0+1=1,\,0+1=1,$
 
 $1+1<n,\,2+1=3,\,w(2)=w(3),\,1+1=2,\,1+1=2,\,\dots,(n-3)+1<n,\,$
 
 $(n-2)+1=n-1,\,w(n-2)=w(n-1),\,(n-3)+1=n-2,\,(n-3)+1=n-2,$
 
 $(n-1)<n,\,(n-1)+1=n,\,w(n-1)\neq{}w(n),\,(n-2)>0,\,(n-3)=n-3,\,w(n-2)\neq{}w(n),$
 
 $(n-3)>0,\,(n-4)=n-4,\,w(n-3)\neq{}w(n),\dots,1>0,\,0=0,\,w(1)\neq{}w(n),$
 
 $0\leq0,\,n\geq{}n,\,0=0,\,r=0$
 
 Replace constants by their values and delete trivially valid literals from the trace literal sequences above. We get  
 
 \NI{}for the trace of $\GA(\sib)$: 
 
 $x(1)=[x(1)],\,x(1)+x(2)=[x(1)+x(2)],\dots,x(1)+\dots+x(n)=[x(1)+\dots+x(n)],$
 
 $\si=[x(1)+\dots+x(n)]$
 
 \VSon{}\NI{}for the trace of $\GA_0(\ff)$:
 
 $w(1)=w(2),\,w(2)=w(3),\,\dots,w(n-2)=w(n-1),\,w(n-1)\neq{}w(n),\dots,w(1)\neq{}w(n),\,\vph=0$
 
  \VSon{}\NI{}for the trace of $\GA_1(\ff)$:
  
  $w(1)=w(2),\,w(2)=w(3),\,\dots,w(n-2)=w(n-1),\,w(n-1)\neq{}w(n),\,w(n-2)\neq{}w(n),$ 
  
  $w(n-3)\neq{}w(n),\dots,w(1)\neq{}w(n),\,r=0$
  
  \VSon{}A \emph{weeded trace} of inputs $X$, denoted $\wtr(X)$, a subsequence of the sequence $(\tl(X,t))_t$ of trace literals  obtained from $(\tl(X,t))_t$ by deleting all constant literals. In a weeded trace, a trace literal that contains the symbol of an input function may be true or not depending on the value of the input, though we consider occurrences of this symbol in trace for a particular input $X$. We leave in $\wtr(X)$  only such non-trivial trace literals. 
  
  We denote by $\wtr(X,k)$ the $k$th element of $\wtr(X)$, and by $\tm(X,\La)$, where $\La\in\wtr(X)$ the time instant $t$ such that $\La=\tl(X,t)$, i.e., such that $\La$ is the trace literal of $\tr(X,t)$.
  
    \VSon{}These `weeded' trace literal sequences are used to estimate entropic convergence below. The literals in these weeded traces  represent events that are directly involved in processing inputs. In the general case one can insert in a `good' algorithm events of this kind that are useless, just to hide what is really necessary to do in order to compute the result. We hope to estimate the usefulness  of events with the help of their entropic weight.

 \section{Inputs partitions and measure}\label{SecInpPartMeasure}

Partitions of $\dm$ are defined by a similarity relation between events that is denoted \mdlr{\sim}. The choice of the probabilistic measure is based on informal \emph{Principle of Maximal Uncertainty}. In examples we use as \mdlr{\sim} the equality of trace literals of events, i.e., two events are similar if their trace literals are equal.  
 
 Let $M=|\rn|$. Fix an order of elements of $\rn=(\om_1,\dots,\om_M)$, and denote 
 $\WDH{\Fn}_k=\Fn^{-1}(\om_k)$. Now the sets $\WDH{\Fn}_k$ are ordered according to $k$.
  
  \VSon{}To an event  \mdlr{E=\tr(X,t)} we relate a set of inputs \mdlr{\WDH{E}=\WDH{E}[X,t]} :
  
  \VSon{}\Hc{2}\mdlr{\WDH{E}[X,t]=\{X’\in\dm:\QE{}t’.\;E\sim\tr(X’,t’)\}} 
  
  \NI(notice, there is no order relation between $t$ and $t'$), 
  
  \NI{}and an ordered partition 

 \VSon\Hc{2}
 $\pe(E)=\pe(\WDH{E})\DF(\WDH{E}\cap{}\WDH{\Fn}_1,\;\WDH{E}\cap{}\WDH{\Fn}_2,\,\dots,\WDH{E}\cap{}\WDH{\Fn}_M)$. 

  \VSon{}In particular, 

 \VSon{}\Hc{2}$\pd\DF\pe(\dm)=(\WDH{\Fn}_1,\;\WDH{\Fn}_2,\;\dots,\;\WDH{\Fn}_M)$

\VSon{}\Hc{2}$\pd_k=\pd_{\Fn_k}\DF\pe(\WDH{\Fn}_k)=(\ES,\;\dots,\ES,\;\WDH{\Fn}_k,\;\ES,\dots,\ES)$ 

\NI{}The latter partition represents the graph of $\Fn$ in our context, we denote it $\gr(\Fn)\DF\{\pd_k\}_k$.

 \VSon{}We define a measure on $\dm$  according to the \emph{Principle of Maximal Uncertainty}. 
 Imagine that $\GA$ plays against an adversary that chooses any input to ensure the maximal uncertainty for 
 $\GA$. 
 In this case all outputs of $\rn(f)$ are equiprobable. We consider a static measure, i.e., that one does not change during the execution of $\GA$. 
  
  We set $\displaystyle\Pr(\WDH{\Fn}_v)=\frac{1}{M}$ for any $v\in\rn(\Fn)$, and define $\Pr$ as uniform on 
  each $\WDH{\Fn}_v$. 
    \VSon{}Practical calculation of $\Pr(S)$ for a set $S$ is combinatorial: 
  $\displaystyle\Pr(S)=\sum_k\,\frac{|S\cap\WDH{\Fn}_k|}{M\cdot|\WDH{\Fn}_k|}$, where where $|S|$ is the cardinality of $S$. The the measure of one point of 
  $\WDH{\Fn}_k$ is $\displaystyle\frac{1}{M\cdot|\WDH{\Fn}_k|}$.  
  
  Remark that we can define a metric between ordered partitions $(A_1,\dots,A_M)$ and $(B_1,\dots,B_M)$: 
 
 \VSon{}\Hc{2.5}$\displaystyle\d((A_1,\dots,A_M),(B_1,\dots,B_M))=
\sum_{1\leq{}i\leq{}M}\Pr(A_i\SD{}B_i)$,

 \VSon{}\NI{}where $\SD$ is symmetric difference of sets, though it remains unclear whether this kind of metric may help to deepen the understanding of algorithmic processes. 
 
  \VSon{}We would like to evaluate the uncertainty of events in a way that says how the algorithm approaches the result. As a measure of uncertainty we introduce a function  $\CD$ over partitions $\pe(E)$ (that can be also seen as a function over events $E$ or sets $\WDH{E}$) that has at least the following properties:
  
   \VSon{}(D1) $\CD(\dm)=\CD(\pd)=\log{}M$ {(maximal uncertainty)},

  \VSon{}(D2) $\CD(\WDH{\Fn}_k)=0$ {(maximal certainty)}, 

  \VSon{}(D3) $\CD(\WDH{E})=0$ for $\WDH{E}\subseteq\WDH{\Fn}_k$ for all $k$, 
  
  \Hc{.8}(the event $E=\tr(X,t)$ determines the result $\Fn(X)$ with certainty), 
 
 \VSon{}(D4) $\CD$ is monotone: it is non-increasing when $\WDH{E}$ diminishes. 
 
 \VStw{} Look at conditional probability $\frac{\Pr(\WDH{E}\cap{}\WDH{\Fn}_k)}{\Pr(\WDH{E})}$. Intuitively, it measures a contribution of event $E$ (via its set $\WDH{E}$) to determining what is the probability to have 
 $\om_k$ as the value of $\Fn$ in trace $\tr(X)$ and in other traces that contain an event similar to $E$.  
 If $\WDH{E}\subseteq\WDH{\Fn}_k$ then 
 $\frac{\Pr(\WDH{E}\cap{}\WDH{\Fn}_k)}{\Pr(\WDH{E})}=1$, i.e., according to $E$ the result is $\om_k$. So we can take as an entropy-like measure this or that average of the conditional information function 
 $-\log\frac{\Pr(\WDH{E}\cap{}\WDH{\Fn}_k)}{\Pr(\WDH{E})}$. As we are interested only in the relation of $E$ with 
 $\Fn$, we take some kind of average over $\WDH{E}$ --- we take it using the measure over $\WDH{E}$ induced by 
 $\Pr$ (then the measure of the whole $\WDH{E}$ may be smaller than $1$). So we  define  
 
  \VSon{}\emph{entropic weight of event}  $E$  (in fact, that of $\pi(E)$) as

          \begin{equation}\label{eqDefEntWght}
 \CD(E)=\CD(\WDH{E})=\CD(\pe(E))=
  -\sum_k{}\Pr(\WDH{E}\cap{}\WDH{\Fn}_k)
   \log\frac{\Pr(\WDH{E}\cap{}\WDH{\Fn}_k)}{\Pr(\WDH{E})}.
          \end{equation}

This function has the properties (D1)--(D4), the properties (D1)--(D3) are evident, and (D4) is proven in Proposition~\ref{propDmonot} below. 

We use in this proof the formula (\ref{eqDefEntWght1}) below  that is equivalent 
to (\ref{eqDefEntWght}) as $\displaystyle\sum_k\,\Pr(\WDH{E}\cap{}\WDH{\Fn}_k)=\Pr(\WDH{E})$:
 
           \begin{equation}\label{eqDefEntWght1}
  \CD(E)=-\sum_k{}\Pr(\WDH{E}\cap{}\WDH{\Fn}_k)
   \log\Pr(\WDH{E}\cap{}\WDH{\Fn}_k) + \Pr(\WDH{E})\log\Pr(\WDH{E})
           \end{equation}
            
  \begin{prop}\label{propDmonot}\PNT
  For any sets $S_0,S_1\subseteq\dm$ if $S_0\subseteq{}S_1$ then $\CD(S_0)\leq\CD(S_1)$   
  \ESqr
  \end{prop}
          
  \NBF{Proof.} Take any function of continuous time $S(t)\subseteq\dm$ such that $S(t_1)\subseteq{}S(t_0)$ for $t_0\leq{}t_1$.   Denote $x_k=\Pr(S(t)\cap{}\WDH{\Fn}_k)$. Then  $\Pr(S(t))=\sum_k\,x_k$. 
  
  We have $0\leq{}x_k\leq\frac{1}{M}$, and 
  $0\leq\sum_k\,x_k\leq1$. We assume that $x_k$ are differentiable. Take derivative of 
  $\CD(S(t))=-\sum_k\,x_k\log{x_k}+(\sum_k\,x_k)\log(\sum_k\,x_k)$ over $t$ (we assume that $S(t)$ is not empty, and for formal reasons we can take only $k$ for which $x_k(t)>0$):
      \begin{eqnarray}
 \CD'(S(t))=-\sum_k\,\Big(x_k'\log{x_k}+x_k\frac{x_k'}{x_k\cdot\ln2}\Big)+
 \Big(\sum_k\,x_k'\Big)\log\Big(\sum_k\,x_k\Big)+
 \Big(\sum_k\,x_k\Big)\frac{\Big(\sum_k\,x_k'\Big)}{\Big(\sum_k\,x_k\Big)\ln2}\nonumber\\
 =-\sum_k\,\Big(x_k'\log{x_k}+\frac{x_k'}{\ln2}\Big)+\Big(\sum_k\,x_k'\log\Big(\sum_k\,x_k\Big) +
 \frac{x_k'}{\ln2}\Big)
 =\sum_k\,x_k'\cdot\log\frac{\Big(\sum_k\,x_k\Big)}{x_k}\label{eqnDerivCD}
      \end{eqnarray}

\begin{figure}[htbp]
  \begin{center}
 \Hc{-1}\includegraphics[width=0.9\textwidth]{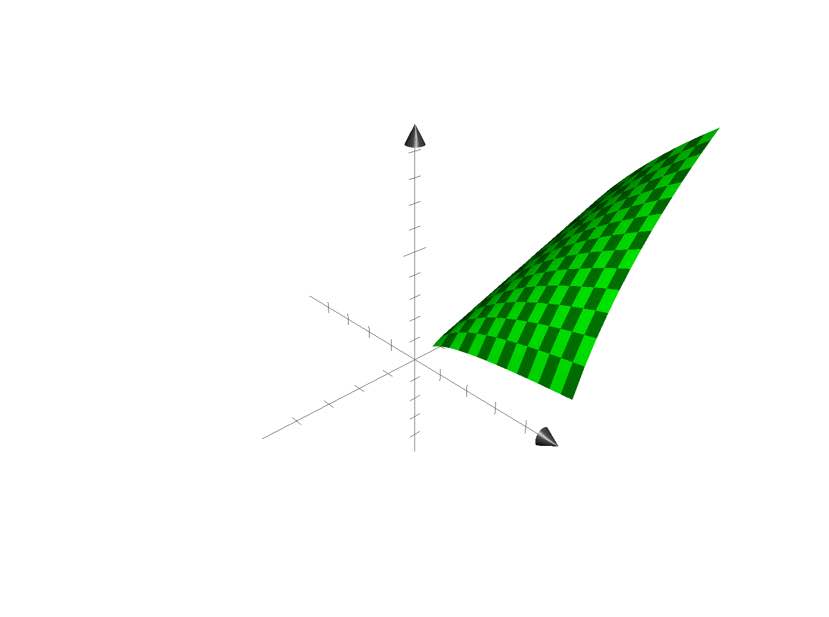}
  \caption{Graph of entropic weight of two variables $z=-x\log{}x-y\log{}y+(x+y)\log(x+y)$}
  \label{FigEntWght2Var}
  \end{center}
       \end{figure}

 The functions $x_k$ are decreasing, thus $x_k'\leq0$.  As $\sum_k\,x_k\geq{}x_k$, the value of (\ref{eqnDerivCD}) is non-positive, hence $\CD(S(t))$ is (non strictly) decreasing when $S(t)$ decreases 
 (see Figure~\ref{FigEntWght2Var}). 
  \Qed
 
\begin{figure}[htbp]
  \begin{center}
 \Hc{-1}\includegraphics[width=0.9\textwidth]{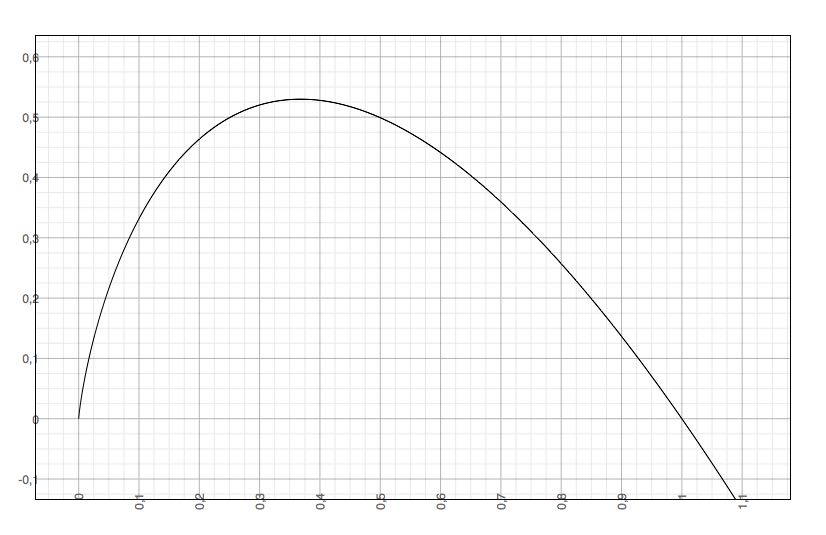}
  \caption{Graph of $y=-x\log{}x$}
  \label{GraphXlogX}
  \end{center}
       \end{figure}

   \begin{prop}\label{propUpBnd}\PNT
  For any $\CJ\subseteq[1..M]$, $M\geq3$, and $S\subseteq\dm$
       \begin{equation}\label{eqUpBndDe}
  \De(S,\CJ)\DF-\sum_{k\in\CJ}\Pr(S\cap{}\WDH{\Fn}_k)
   \log\frac{\Pr(S\cap{}\WDH{\Fn}_k)}{\Pr(S)}\leq  \frac{|\CJ|}{M}\log{}M
     \end{equation}
  \ESqr
  \end{prop}
\NBF{Proof.}We have
     \begin{eqnarray}
  \De(S,\CJ)\DF-\sum_{k\in\CJ}\Pr(S\cap{}\WDH{\Fn}_k)
   \log\frac{\Pr(S\cap{}\WDH{\Fn}_k)}{\Pr(S)}\leq 
   -\sum_{k\in\CJ}\Pr(S\cap{}\WDH{\Fn}_k)
   \log\frac{\Pr(S\cap{}\WDH{\Fn}_k)}{\Pr(\dm)}=\nonumber\\
   -\sum_{k\in\CJ}\Pr(S\cap{}\WDH{\Fn}_k)
   \log\Pr(S\cap{}\WDH{\Fn}_k)
     \end{eqnarray}
Function $x\log{}x$ is increasing for $0\leq{}x\leq0.36<\frac{1}{e}$, where $e$ is the base of natural logarithm (see Figure~\ref{GraphXlogX}). 

Indeed, take derivative of   $-x\log{}x=-\frac{1}{\ln2}x\ln{}x$. We get 
$-\frac{1}{\ln2}(\ln{}x+1)$; this expression is zero when $\ln{}x=-1$, i.e., $x=\frac{1}{e}$. And the derivative is positive for 
$0\leq{}x<\frac{1}{e}$. 

Thus, for $M\geq3$ the right-hand side of (\ref{eqUpBndDe})  is a sum of  functions increasing for 
$0\leq\Pr(S\cap{}\WDH{\Fn}_k)\leq\frac{1}{3}$ when $S$ grows. Hence, 
    \begin{eqnarray}
  \De(S,\CJ)\leq-\sum_{k\in\CJ}\Pr(\WDH{\Fn}_k)\log\Pr(\WDH{\Fn}_k)=
  -\sum_{k\in\CJ}\frac{1}{M}\log\frac{1}{M}=
  \frac{|\CJ|}{M}\log{}M,
     \end{eqnarray}
that gives (\ref{eqUpBndDe}). \Qed

   \begin{prop}\label{propUpBndSpc}\PNT
  For any $\CJ\subseteq[1..M]$  and $S\subseteq\dm$ such that $S\cap\WDH{\Fn}_k=\ES$ for all $k\notin\CJ$, there holds
  \begin{equation}\label{eqDeSparce}
  \De(S,\CJ)\leq\Pr(S)\log|\CJ|\leq\log|\CJ|
   \end{equation}
where we use notation from (\ref{eqUpBndDe}). 
   \ESqr
  \end{prop}
\NBF{Proof.} Clearly, 
$\Pr(S)=\sum_{1\leq{}k\leq{}M}\Pr(S\cap\WDH{\Fn}_k)=\sum_{k\in\CJ}\Pr(S\cap\WDH{\Fn}_k)$,  
$\sum_{k\in\CJ}\frac{\Pr(S\cap\WDH{\Fn}_k)}{\Pr(S)}=1$, hence, $\frac{\Pr(S\cap\WDH{\Fn}_k)}{\Pr(S)}$ is a probability distribution, and the maximal value of its entropy is
  \begin{eqnarray}
 -\sum_{k\in\CJ}\frac{\Pr(S\cap\WDH{\Fn}_k)}{\Pr(S)}\log\frac{\Pr(S\cap\WDH{\Fn}_k)}{\Pr(S)}\leq
 \log|\CJ| \MEQ \De(S,\CJ)\leq\Pr(S)\log|\CJ|.\label{eqDeSparcePrf}
  \end{eqnarray}
As $\Pr(S)\leq1$ from (\ref{eqDeSparcePrf}) we get (\ref{eqDeSparce}). 
 \Qed

Clearly, the bound of Proposition~\ref{propUpBnd}, when applicable, is better than the last inequality of Proposiiton~\ref{propUpBndSpc} except one small value of $|\CJ|$. In our applications $|\CJ|=(M-k)$ with $k$ going from $0$ to $M$. Thus, for the upper bounds of the mentioned Propositions we have $\frac{M-k}{M}\log{}M\leq\log(M-k)$ for $k<(M-1)$.  

 In order to understand entropic convergence of $\GA$ we can look at the behavior of the entropic weight along individual traces, mainly corresponding to the worst-case complexity, or to look at the evolution of the entropic weights of the set of all events after a given time instant that goes to $\infty$. Some events, e.g., related to the updates of loop counters, may be not really related to the convergence of $\GA$ to the result, and hence, should not be taken into consideration because of evident reasons that are commented in the examples of 
 Section~\ref{SecAnlzExmp}.  However, the choice of relevant events is not governed by a rigorous formal procedure, at least at the present stage of study. What is relevant and what not is clear in concrete situations, however, one can imagine algorithms where `the relevance' is well hidden artificially.

  \section{Analysis of examples}\label{SecAnlzExmp}
  
  Here we take as similarity relation $\sim$ the equality of trace literals, i.e., 
  $E\sim{}E'$ if $\tl(E)=\tl(E')$. In order to have a point of departure we tacitly always take into consideration the first step of initialisation in $\GA$, and notice that the entropic weight of this event is maximal, i.e., $\log{}M$. 
  
  \VStw{}\NBF{Example~\ref{exmpSumMod2}.  $\GA(\sib)$: sum over $\F_2$. Convergence.}
  
  \VSon{}\NI{}Trace literals of $\GA(\sib)$ are of the form $1_k=k$, where $1_k$ is an expression $0+1+\dots+1$ containing $k$ symbols of the constant $1$, or of the form $\si_k(x)=[x(1)]+\dots+[x(k)]$, where 
  $\si_k(x)=x(1)+\dots+x(k)$. For any event $E$ that represents an update of loop counter $i$ we have $\WDH{E}=\dm$, and thus, $\CD(E)$ has its maximal value, and hence says nothing about the convergence of $\GA(\sib)$ to the result. We do not take these events into consideration. We can exclude them using a general `filter': throw away all events whose trace literal is trivially true, i.e., is true whatever be inputs (if the literal contains ones). E.g., this filter eliminates literals like $x(i)=x(i)$ or $x(i)+x(i)=0$. We call the remaining events of the form $\si_k=a$, where $a\in\B$, \emph{essential}. Notice that this notion of essential works well for our examples; in the general case the analysis of convergence is more complicated.
  
  Events of with trace literal $\si_k(x)=a$ take place at instants $2+3k$. Denote the set 
  $\{x:\si_k(x)=a\}$ by $(\si_k=a)$, and the set $(\si_k=a)\cap(\si_n=b)$ by $S_{k,a,b}$, where $a,b\in\B$. Notice, that $\si_n=\si$.
  
   For any $a\in\B$ we have $\CD(\si_k=a)=\frac{1}{2^k}$. 
  Indeed, 
  
  \VSon{}$\Pr(\si_k=a)=\frac{2^{n-k}}{2^n}=\frac{1}{2^k}$, 
  
  \VSon{}$\Pr(S_{k,a,0})=\Pr(S_{k,a,1})=\frac{1}{2^{k+1}}$, 
  
  \VSon{}$\CD(\si_k=a)=-\Pr(S_{k,a,0})\log\frac{\Pr(S_{k,a,0})}{\Pr(\si_k=a)}- 
   \Pr(S_{k,a,1})\log\frac{\Pr(S_{k,a,1})}{\Pr(\si_k=a)}=
  -\frac{1}{2^{k+1}}\log\frac{1}{2}-\frac{1}{2^{k+1}}\log\frac{1}{2}=\frac{1}{2^k}$.

  \VSon{}This describes the convergence along traces: $\CD(\tr(x,t))=\frac{1}{2^k}$, where $k=\frac{t-2}{3}$. 
  
  \VSon{}Look at the space $\GR(t)$ consisting of all essential events that happen at $t$ or later. 
  We have $2^{k}$ 
  events $\si_k=a$ with $\CD(\si_k=a)=\frac{1}{2^k}$. We evaluate the weighted volume of $\GR(t)$, i.e.,  the volume where for each element we take its entropic weight. Denote this volume  $\CD\GR(t)$.
  
   We have 
  
  \[\CD\GR(2+3k)=\sum_{j=k}^n\,2^{j}\cdot\frac{1}{2^j}=n-k+1,\;\; \textrm{or}\;  \CD\GR(t)=n-\frac{t+1}{3}.\]
  
  We see that in terms of weighted volume the convergence is linear and monotone; the convergence along traces is also monotone. Thai is not the case for $\maxPS$ considered below.
  
  \VStw{}\NBF{Example~\ref{exmpMaxPS}. Maximal prefix-suffix ($\maxPS$): $\vphb$. Convergence.}

\VSon{}\NI{}For algorithms $\GA_m(\vphb)$ the events whose trace literals are constant play a non-trivial role in understanding the convergence, as compared to the case of $\GA(\sib)$, but however, we apply the same filter as for $\GA(\sib)$ to define essential events because the essential ones suffice to estimate the convergence. 
So essential events are equalities and inequalities of input characters $w(i)$, where $i$ a concrete natural number. (One can notice that these events contain the information given by constant literals decsribing loops.)

We order $\rn(\vphb)=(0,\dots,(n-1))$. 

 The measure $\Pr$ for this problem is hard to calculate exactly (as far as I am aware, even estimation of
 $|\Hph_k|$ for $k<\frac{n}{2}$ is an open problem):
          \begin{equation}\label{eqMesFormaxPS}
\Pr(S)=\sum_{k=0}^{n-1}\,\Pr(S\cap\Hph_k)=\sum_{k=0}^{n-1}\frac{|S\cap\Hph_k|}{n\cdot|\Hph_k|},
           \end{equation}
but we wish to understand very approximately the behavior of $\CD(X,t)$ as a function of $t$ for the worst-case 
inputs $w$.  For $\GA_m(\vphb)$ a worst-case input is, e.g., $w_1=aa\dots{}ab$ with $a\neq{}b$.

 \VSon{}\NBF{Algorithm $\GA_0(\vphb)$}. We consider the work of $\GA_0(\vphb)$ over $w_1$.
 For further references we introduce notations $\xi_i$ for pieces of  $\wtr(w_1)$):
  
       \begin{equation}\label{eqmaxPS_DSzero}
   \left.
 \begin{array}{ll}
 \xi_{n-1}:\;w(1)=w(2),\;w(2)=w(3),\;\dots,\;{}w(n-1)\neq{}w(n)\\
 \xi_{n-2}:\;w(1)=w(3),\;{}w(2)=w(4),\;\dots,\;{}w(n-2)\neq{}w(n)\\
   \dotfill\\
 \xi_{2}:\;w(1)=w(n-1),\;{}w(2)\neq{}w(n)\\
 \xi_{1}:\;w(1)\neq{}w(n)
  \end{array}
 \right\}  
      \end{equation}
\Hc{3}$\xi_{k,i}\DF{}\big(w(i)=w(i+(n-k))\big)$ \ \ for $1\leq{}i\leq(k-1)$, \;\;
 $\xi_{k,k}\DF{}\big(w(k)\neq{}w(n)\big)$.  
 
 \VStw{}One can see that entropic weight of the last event of $\xi_{n-1}$ is zero: $\CD(\xi_{n-1,n-1})=0$. Indeed, 
 
 $\WDH{\xi}_{n-1,n-1}=\{w:\xi_{n-1,1}\wedge\dots\wedge\xi_{n-1,n-2}\wedge\xi_{n-1,n-1}\}=
  \{a\dots{}ab: a\neq{}b\}\subset\Hph_0$, \ \ and  $\CD(\xi_{n-1,n-1})=0$ due to (D3).
  
  \NI{}Thus, after event $\xi_{n-1,n-1}$ algorithm $\GA_0(\vphb)$ has enough information to decide what is the result, however it continues to work. We try to look at what goes on before and after this event.
  
    \VSon{}Set $S=\WDH{\xi}_{n-1,s}=\{w: w(1)=w(2)\wedge\dots\wedge{}w(s)=w(s+1)\}$.

 \VSon{} It is evident that \VSon$S\cap\Hph_{n-1}=\Hph_{n-1}$. 
 \begin{lem}\PNT
$S\cap\Hph_{i}=\ES$ for $(n-s-1)\leq{}i\leq(n-2)$.
  \end{lem}
\NBF{Proof.} Suppose there is $w\in(S\cap\Hph_{i})$. Then $w\in{}S$ implies that $w=a^{s+1}v$, where $a$ is a character and $v$ is a word of length $(n-s-1)$. From $w\in\Hph_{i}$ we see that $w(1..i)=w((n-i+1)..n)$. 

    The inequality $(n-s-1)\leq{}i$ gives $n-i+1\leq{}n-(n-s-1)+1=s+2$ that means that the word $w((n-i+1)..n)$ either immediately follows $a^{s+1}$ or even intersects it.  
 As $i\geq(n-s-1)$  the first $(n-s-1)$ characters  of  $w((n-i+1)..n)$ coincide with $w(1..i)$. 
Two cases are possible.

 Case~1: $i\leq(s+1)$. Then $w(1..i)=a^i=w((n-i+1)..n)$. Hence, $w=a^n$ and $\vphb(w)=n-1$ that is excluded by the premise of Lemma.
 
 Case~2: $i>(s+1)$. As $n-i+1\leq{}s+2\leq{}i$, then $w(1..i)$ and $w((n-i+1)..n)$ intersect, and thus $w$ is periodic with a period of length $n-i+1$. But as $n-i+1\leq{}s+2$, this period has a form $a^{n-i+1}$, and hence, again $w=a^n$ and $\vphb(w)=n-1$ that is excluded by the premise of Lemma.
\Qed   

 Proposition~\ref{propUpBnd} give a linear upper bound $(1-\frac{k}{n})\log{}n$ on the speed of convergence the 
  the entropic weight of $\xi_{n-1,n-1}$: $\CD(\xi_{n-1,n-1})=0$.
  
 Event $\xi_{n-2,1}$  immediately follows event $\xi_{n-1,n-1}$.  Denote $G\DF\WDH{\xi}_{n-2,1}$. It is intuitively clear that the entropic weight of $G$ is rather big. We give a weak estimation that is qualitatively sufficient to make such a conclusion, and thus, for the analysis of the behavior of $\GA_0(\vph)$:
 
     \begin{equation}\label{eqLowBndXi_n-2_1}
 \CD(G)\geq\frac{\log{}n}{4\al}-c(\al). 
     \end{equation}
The (boring and not so instructive) calculations that give this bound are in Annexe, 
subsection~\ref{subSecLwBndXi_n-2_1}.

 \VSon{}We see that after event $\xi_{n-1,n-1}$ with entropic weight zero, $\GA_0(\vphb)$ executes an event whose entropic weight jumps up to at least $\frac{\log{}n}{4\al}-c(\al)$.  After that the entropic weight  goes down  to   $\CD(\xi_{n-2,n-2})$ that we show just below.  
 
 The event $\xi_{n-2,n-2}$ can happen only for words of the form $w_1\DF{}ab\dots{}abac$ 
 with $a\neq{}b$ and $c\neq{}b$ (if $n$ is even), or of the form $w_2\DF{}ab\dots{}abc$ with $a\neq{}b$ and 
 $c\neq{}a$ (if $n$ is odd). If $c\neq{}a$ in $w_1$ then $\vphb(w_1)=0$, and if $c=a$ then $\vphb(w_1)=1$. As for $w_2$ it is always $\vphb(w_2)=0$. 
 
 \VSon{}Thus $\CD(\xi_{n-2,n-2})=0$ for odd $n$. 
 
 Let $n$ be even. Denote 
 $W_0\DF\{ab\dots{}abac: a\neq{}b\wedge\,c\neq{}a\}$, $W_1\DF\{ab\dots{}abaa: a\neq{}b\}$. Clearly, 
 $|W_0|=\al(\al-1)(\al-2)$, $|W_1|=\al(\al-1)$. Let $H=W_0\cup{}W_1$, with this notation 
 $H=\WDH{\xi}_{n-2,n-2}$. As it was mentioned just above $H\cap\Hph_0=W_0$ and $H\cap\Hph_1=W_1$.
 
 We can show that $\CD(H)$ is `small' (see Annexe, subsection~\ref{UpBndXi_n-2_n-2}): 
 
  \begin{equation}\label{eqUpBndXi_n-2_n-2}
 \CD(H)\leq\BO{\frac{1}{\al^{\frac{n}{2}-3}}}
  \end{equation}
 
 We see that $\CD(H)=\CD(\xi_{n-2,n-2})$ is either zero or very small. We observe that at this point the behavior of $\GA_0(\vphb)$ is irregular, and though later these irregularities diminish, however, in order to eliminate a value $k$ of $\vphb$ the algorithm makes $k$ comparisons of characters. The general convergence can be estimated as follows.
 
 After event $\xi_{k,k}$ the value $k$ of $\vphb$ is eliminated, as well as all bigger ones. The entropic weight of $\xi_{k,k}$ grows down as a function of $k$, and the speed of this convergence is given by 
 Proposition~\ref{propUpBnd}. We see this convergence takes much of time, namely, in order to arrive at $\xi_{k,k}$ the algorithm $\GA_0(\vphb)$ makes about $(n-k)^2$ steps. And we see also that the entropic weight behaves irregularly, not smoothly, namely, it goes up and down. All this shows that the extraction of information of $\GA_0(\vphb)$ is not efficient.
  
 \VStw{}\NBF{Algorithm $\GA_1(\vphb)$}. We see that for $\GA_1(\vphb)$, as it was for $\GA_0(\vphb)$, the entropic weight of event $w(n-1)\neq{}w(n)$ is zero. The next event is $w(n-2)\neq{}w(n)$. Its entropic weight can be evaluated as above, and it is `very small'. And one value of $\vphb$ is eliminated as in the case of 
 $\GA_0(\vphb)$. The next event is $w(n-3)\neq{}w(n)$. It may happen only for words of the form $(abc)^ma'$ or 
 $(abc)^mab'$ or $(abc)^mabc'$ with respectively $a\neq{}a'$, $b\neq{}b'$ and $c\neq{}c'$. Though possible values of $\vphb$ for such words are $0,1,2$, their measure is small though slightly bigger that in the previous case
  that is given by (\ref{eqUpBndXi_n-2_n-2}). But this event eliminates the value $(n-3)$ of $\vphb$. The latter is in some way more important. Each next inequality $w(k)\neq{}w(n)$ again eliminates a value of $\vphb$, and we can again apply   Proposition~\ref{propUpBnd} to estimate the convergence. On the whole we can see a small increasing of entropic weight up to some point after which  the eliminated values start to ensure  the convergence of entropic weight to zero. 
 
\VStw{}\NBF{Remark.} We can explain a similar convergence of $\GA_0(\vphb)$ and$\GA_1(\vphb)$ in  `purely logical' way that does not refer to the graph of $\vphb$, and for this reason cannot be extended to problems independently 
of algorithms.

  Formulas (\ref{eqmaxPS_DSzero}) are in fact trace formulas as defined just below.
      
     \VSon{}\emph{Trace formula} is a formula like $\TF(X,t)\DF\bigwedge_{1\leq{}\ta\leq{}t}\tl(X,\ta)$ or 
     $\TF(X)\DF\TF(X,\infty)$. 
   
   (Notice that the updates of loop counters give trivial tautologies.)
   
   \VSon{}\emph{Trace subformula} is a formula of the form $\TF(X,\CI)\DF\bigwedge_{\ta\in\CI}\tl(X,\ta)$, where 
   $\CI\subseteq[1..\infty]$.
   
   \VSse{}Notation: $\WDH{\Ps}\DF\{X:\Ps(X)\}$
   
   \VSon{}A formula $\Ps$ is a \emph{defining formula (DF)} of $\Fn(X)$ if 
   $\WDH{\Ps}\subseteq{}\Fn^{-1}(\Fn(X))$. A formula $\Ps$ is a \emph{minimal defining formula (MDF)} of 
   $\Fn(X)$ if $\Ps$ is DF of $\Fn(X)$, and no its subformula is DF of $\Fn(X)$.
  
 \VSon{}Any $\GA_k(\vphb)$ constructs its own defining formula for the result. So we can estimate its convergence `towards this defining formula'. 
 For words in $z\in\vphb^{-1}(0)$ the formula used by $\GA_0(\vphb)$ is 
 
         \begin{equation}\label{eqmaxPS-DSzero}
   \left.
 \begin{array}{ll}
 \ze_{n-1}:\;z(1)\neq{}z(2)\vee{}z(2)\neq{}z(3)\vee\dots\vee{}z(n-1)\neq{}z(n)\\
 \ze_{n-2}:\;z(1)\neq{}z(3)\vee{}z(2)\neq{}z(4)\vee\dots\vee{}z(n-2)\neq{}z(n)\\
   \dotfill\\
\ze_{2}:\;z(1)\neq{}z(n-1)\vee{}z(2)\neq{}z(n)\\
\ze_{1}:\;z(1)\neq{}z(n)
  \end{array}
 \right\}  
      \end{equation}
For the general case $z\in\vphb^{-1}(v)$ we have

         \begin{equation}\label{eqmaxPS-DS}
   \left.
 \begin{array}{ll}
 \ze_{n-1}:\;z(1)\neq{}z(2)\vee{}z(2)\neq{}z(3)\vee\dots\vee{}z(n-1)\neq{}z(n)\\
   \dotfill\\
 \ze_{v+1}:\;z(1)\neq{}z(v+1)\vee{}z(2)\neq{}z(v+2)\vee\dots\vee{}z(n-v)\neq{}z(n)\\
 \OVL{\ze}_{v}(\MEQ\neg\ze_v):\;z(1..v)=z(n-v+1..n)
  \end{array}
 \right\}  
      \end{equation}
      
  In order to find the value of $\vphb(z)$ for an input $z$ (suppose  $\vphb(z)=v$) algorithm 
  $\GA_0(\vphb)$  constructs a sequence of inequalities, at least one from each  $\ze_{i}$, 
  $(v+1)\leq{}i\leq(n-1)$, and the equalities $\OVL{\ze}_{v}$. 
  
  Look at the convergence of $\GA_k(\vphb)$ towards the defining formula for input $w=a^{n-1}b$ from the  viewpoint of the Principle of Maximal Uncertainty.  We try to choose a model that gives an intuitively acceptable explanation of the convergence (other models are also imaginable). 
  
  Any $\GA_k(\vphb)$, $k=0,1$, starts its work with the verification of $\ze_{n-1}$ from left to right. As all the values of $\vphb$ are equiprobable (then the uncertainty of final result is maximal), the probability of 
  $\neg\ze_{n-1}$ is $\frac{1}{n}$, and that of $\ze_{n-1}$ is $(1-\frac{1}{n})=\frac{n-1}{n}$.
  
  For the same reason of maximizing the uncertainty, the probability to have $z(1)\neq{}z(2)$ \ is \ 
  $\frac{n-1}{n(n-1)}=\frac{1}{n}$. If $z(1)={}z(2)$ then probability to have $z(2)\neq{}z(3)$ becomes slightly bigger: $\frac{n-1}{n(n-2)}=\frac{1}{n}+\frac{1}{n(n-2)}$. And so on: if 
  $\bigwedge_{i\leq{}p}z(i-1)={}z(i)$ then the probability to have $z(j)\neq{}z(j+1)$  for 
  $j\geq{}p$ is 
  $\frac{n-1}{n(n-p)}=\frac{1}{n}+\frac{p-1}{n(n-p)}$. 
  
  \vspace{1ex}
   After $\ze_{n-1}$ has been established, the value $\vphb(z)=n-1$ is excluded, the number of remaining values 
   $\vphb(z)$ becomes $(n-1)$, and the probability of each becomes $\frac{1}{n-1}$. 

   \vspace{1ex}
  After that $\GA_0(\vphb)$ and $\GA_1(\vphb)$ work differently. 
  
  Algorithm $\GA_0(\vphb)$ consecutively checks all $\ze_v$, starting with  $\ze_{n-1}$ . During this processing,  after  $\bigwedge_{i\leq{}p}z(i-1)={}z(i+n-s)$ has been established, the probability to have 
$z(j)\neq{}z(j+n-s)$ for $j\geq{}p$ is $\displaystyle\frac{s-1}{s(s-p)}$, and there are $(s-p)$ such possibilities.

  Thus the entropy of this distribution is 
  \begin{eqnarray}
 -\Big((s-p)\cdot\frac{s-1}{s(s-p)}\log\frac{s-1}{s(s-p)}+\frac{1}{s}\log\frac{1}{s}\Big)=\nonumber\\
 -\left(\left(1-\frac{1}{s}\right)\log\frac{1}{s-p}+
 \left(1-\frac{1}{s}\right)\log\left(1-\frac{1}{s}\right)+
   \frac{1}{s}\log\frac{1}{s} \right)=\nonumber\\
       \log{}s-\frac{\log{}s}{s}+
       \left(1-\frac{1}{s}\right)\log\left(1-\frac{p}{s}\right)-
    \left(1-\frac{1}{s}\right)\log\left(1-\frac{1}{s}\right)+
   \frac{\log{}s}{s}=\nonumber\\
   \log{s}+
       \left(1-\frac{1}{s}\right)\log\left(1-\frac{p}{s}\right)-
    \left(1-\frac{1}{s}\right)\log\left(1-\frac{1}{s}\right)=\nonumber\\
    \log{s}-\left(1-\frac{1}{s}\right)\log\frac{s-1}{s-p}
  \end{eqnarray}
 Here $p\IM(s-1)$ and $s\IM1$ give the speed of diminishing of the uncertainty in terms of this evolution of $s$ and $p$. The convergence by $p$ is very slow and `explains' the complexity 
 $\BO{n^2}$ of $\GA_0(\vphb)$.
 
 The convergence of $\GA_1(\vphb)$ is the same as that of $\GA_0(\vphb)$ only when $\GA_1(\vphb)$ processes 
 $\ze_{n-1}$.  After that there is no $p$, algorithm $\GA_1(\vphb)$ excludes one value of $\vphb$ at each step (that consists of the calculation of $\vphb^{(n-s+1)}(n-1)$ from $\vphb^{(n-s)}(n-1)$ and of the comparison of the appropriate characters), and the  uncertainty goes down only due to $s$, thus much faster. We omit technical details. 
  
   \section {Conclusion}
   
   This text shows that it is not impossible to evaluate algorithmic processes from entropic viewpoint. This is a modest step in this direction. There may be be other approaches, other entropy-style functions or metrics that can play a similar role.
   
   The combinatoric that arises in the present setting is very complicated. Some people may treat this as a shortcoming, the others as a stimulus to develop new  methods for solving combinatorial problems.
   
   One visible constraint of the method is that the number of partitions of inputs is limited by an exponential function of the domain $\dm_n(\Fn)$. However, I think this is not a real constraint. For problems in 
   $\bmth{NP}\cup{}\bmth{coNP}$ whose domains are of exponential size there are enough of partitions. As for problems of higher complexity classes, they are not of the same structure, and their inputs code, in fact, longer inputs.
   
The main challenge is to extend such approaches to algorithmic problems. It seems possible.     

\VStw{}\NBF{Acknowledgements} I am thankful to Eugène~Asarine, Vladimir~Lifschitz and Laurent~Bienvenu for discussions and comments that were useful for me.  

 \section{Annexe: estimations of entropic weight related to maxPS}
 
 Trivial relations:
 
 $\Pr(S\cap\Hph_k)\leq\Pr(\Hph_k)=\frac{1}{n}$,  
 
$\th<\th'$ $\MEQ$ $\log\th<\log\th'$, $-\log\th'<-\log\th$, $\log\frac{1}{\th'}<\log\frac{1}{\th}$, 
     $-\log\frac{1}{\th}<-\log\frac{1}{\th'}$

 \subsection{Lower bound for $\CD(\xi_{n-2,1})=\CD(G)=\CD(\xi_{n-2,1})$.}\label{subSecLwBndXi_n-2_1}
 
  Recall that $G$ is $\xi_{n-2,1}$, i.e., event $w(1)=w(3)$. Clearly, $G\cap\Hph_{n-1}=\ES$ and 
  $G\cap\Hph_{n-2}=\Hph_{n-2}$. Denote $G'\DF\{w:w(1)=w(3)\}$. This set contains $\Hph_{n-1}$ contrary to $G$. Nevertheless for $k\leq(n-2)$ we have $(G\cap\Hph_k)=(G'\cap\Hph_k)$ as for $k\leq(n-2)$ the equality $w(1)=w(3)$ is the only constraint to take into account for $G$. 
 
 Any set $\Hph_k$ with $(n-1)\geq{}k\geq\frac{n}{2}$  consists of periodic words with periods of length $(n-k)$, and each such period is primitive, i.e., cannot be represented as $u^i$ with non-empty $u$ and $i\geq2$ (otherwise, the word has a smaller period and thus, belongs to $\Hph_k$ with bigger $k$). Hence, $|\Hph_{n-s}|$ is equal to the number of primary words of length $s$. 
 The sets $(G\cap\Hph_k)$, $\frac{n}{2}\leq{}k\leq(n-3)$, are also of the same type but whose periods are chosen from the words satisfying  $w(1)=w(3)$.
 
 \VSon{}Let $1\leq{}s\leq\frac{n}{2}$. Denote: $\ga(s)\DF|\Hph_{n-s}|$ (the number of periodic words with primary period of length $s$); and   $\Ga(s)\DF|\Hph_{n-s}\cap{}G'|$ (the number of periodic word with primary period of length  $s$ and such that $G'$). As it was noticed just above, $\Ga(s)=|\Hph_{n-s}\cap{}G|$ for 
 $2\leq{}s\leq\frac{n}{2}$.

 A known formula for $\ga$ is 
 
      \begin{equation}\label{eqMobiusga}
  \ga(s)=\sum_{1\leq{}d\leq{}s,d|s}\,\al^d\mu\Big(\frac{s}{d}\Big)=
  \al^{s}+\sum_{1\leq{}d<s,d|s}\,\al^d\mu\Big(\frac{s}{d}\Big),
      \end{equation}
where $\mu$ is  Möbius function: $\mu(m)=0$ if $m$ is divisible by a square different from $1$, $\mu(m)=(-1)^r$ if $m$ is not divisible by a square different from $1$ and $r$ is the number of prime divisors of $m$; 
$\mu(1)=1$.

 It follows from (\ref{eqMobiusga}) or is easy to verify directly  
   \begin{eqnarray}\label{eqGa_gaSmallArg}
 \ga(1)=\Ga(1)=\al, \;\ga(2)=\Ga(2)=\al^2-\al, \;\ga(3)=\al^3-\al,\;\Ga(3)=\al^2-\al=\Ga(2)=\ga(2), \nonumber\\
 \ga(4)=\al^4-\al^2,\;\Ga(4)=\al^3-\al^2,\;\ga(5)=\al^5-\al,\;\Ga(5)=\al^4-\al
    \end{eqnarray}

   We can calculate $\ga(s)$, as well as $\Ga(s)$, as follows. The number of all words of length $s$ is $\al^s$, and the number of all words of length $s\geq3$ such that $G'$, is $\al^{s-1}$. From these words we subtract the words that are not primary, this can be defined recursively: 

    \begin{eqnarray}\label{eqGaEtga}
 \ga(s)=\al^{s}-\sum_{1\leq{}d<s,d|s}\,\ga(d), \Hc{1}
 \Ga(s)=\al^{s-1}-\sum_{1\leq{}d<s,d|s}\,\Ga(d)
    \end{eqnarray}
 
 Comparing the formulas (\ref{eqMobiusga}) and (\ref{eqGaEtga}) we see that $\ga(s)$ and $\Ga(s)$ can be expressed in terms of powers $\al^p$ with coefficient $1$. So if such an expression does not contain $\al$ or 
 $\al^2$ (that are related to $s=1,2$) then $\Ga(s)=\frac{1}{\al}\ga(s)$. But because of possible presence of  
 $\al$ or $\al^2$, and formulas for  $\Ga(1)$, $\Ga(2)$ and $\Ga(2)$ above, we can only state  that 
    \begin{equation}\label{eqBndsGa_gaSmallVal}
\frac{1}{\al}\ga(s)-\al^2-\al\leq\Ga(s)\leq\frac{1}{\al}\ga(s)+\al^2+\al
     \end{equation}

  For a lower bound we notice that the biggest diviser of $s$ that is smaller than $s$ is not greater than $\frac{s}{2}$, thus 
 
 $\ga(s)\geq\al^s-(\al^{\frac{s}{2}}+\dots+\al)=\al^s-\al\frac{\al^{\frac{s}{2}}-1}{\al-1}>
 \al^s-\al^{\frac{s}{2}+1}$ as $\al\frac{\al^{\frac{s}{2}}-1}{\al-1}<\al^{\frac{s}{2}+1}$, the latter is equivalent to $\al^{\frac{s}{2}}-1<\al^{\frac{s}{2}}(\al-1)\MEQ2\al^{\frac{s}{2}}<\al^{\frac{s}{2}+1}+1$, it rests to notice that $2\leq\al$.
 
 \VSon{}We summarize these these inequalities in 
 
    \begin{equation}\label{eqLowBndGaga}
   \frac{\Ga(s)}{\ga(s)} \geq \frac{1}{\al}-\frac{\al(\al+1)}{\ga(s)},\Hc{1}
   \ga(s) \geq \al^s-\al^{\frac{s}{2}+1}, \Hc{1}\frac{\al^m-1}{\al-1} \leq \al^m.
    \end{equation}
From (\ref{eqLowBndGaga}) and $\displaystyle\al^{s-3}-\al^{\frac{s}{2}-2}\geq
\al^{\frac{s}{2}-2}(\al^{\frac{s}{2}-1}-1)\geq2\cdot(4-1)=6$ for $s\geq6$ we get (for $s\geq6$)

         \begin{eqnarray}\label{eqAlpha_ga}
 \frac{\al(\al+1)}{\ga(s)} \leq \frac{\al(\al+1)}{\al^s-\al^{\frac{s}{2}+1}} \leq  
 \frac{\frac{3}{2}\al^2}{\al^s-\al^{\frac{s}{2}+1}} = \frac{3}{2(\al^{s-2}-\al^{\frac{s}{2}-1})} = 
 \frac{3}{2\al(\al^{s-3}-\al^{\frac{s}{2}-2})}\leq\frac{1}{4\al}      
           \end{eqnarray}
From (\ref{eqLowBndGaga}) and (\ref{eqAlpha_ga}) for $6\leq{}s\leq\frac{n}{2}$ we have
     
        \begin{eqnarray}\label{eqGa_ga_6+}
  \frac{\Ga(s)}{\ga(s)} \geq \frac{1}{\al}-\frac{\al(\al+1)}{\al^s-\al^{\frac{s}{2}+1}}\geq 
  \frac{1}{\al}- \frac{1}{4\al} = \frac{3}{4\al}
        \end{eqnarray}

  From formulas above for $\Ga(s)$ and $\ga(s)$ for $s=3,4,5$ we see that 
  
  \begin{equation}\label{eqGa_ga3_4_5}
 \frac{\Ga(3)}{\ga(3)}=\frac{\al^2-\al}{\al^3-\al}=\frac{1}{\al+1}\geq\frac{1}{2\al},\;\;
 \frac{\Ga(4)}{\ga(4)}=\frac{\al^3-\al^2}{\al^4-\al^2}=\frac{1}{\al+1}\geq\frac{1}{2\al},\;\;
 \frac{\Ga(5)}{\ga(5)}=\frac{\al^4-\al}{\al^5-\al}=\frac{\al^3-1}{\al^4-1}\geq\frac{3}{4\al}
   \end{equation}

From (\ref{eqGa_ga_6+}) and (\ref{eqGa_ga3_4_5}) for $s\geq3$

  \begin{equation}\label{eqLwBndPrGcap_vph}
 \Pr(G\cap\Hph_k)=\frac{\Ga(n-k)}{n\ga(n-k)}\geq\frac{1}{n\cdot2\al}
   \end{equation}

 Recall 
      \begin{eqnarray}\label{eqPrForGvph}
 \Pr(G)=\sum_k\,\Pr(G\cap\Hph_k)>\frac{1}{n}+\sum_{k=n-3}^{\frac{n}{2}}\,\Pr(G\cap\Hph_k)
     \end{eqnarray}

  From  (\ref{eqPrForGvph}), (\ref{eqGa_ga_6+}) and (\ref{eqGa_ga3_4_5}) and the remark on the relation of $G$ and $G'$ we conclude that
  
      \begin{eqnarray}
\Pr(G)=\sum_k\,\Pr(G\cap\Hph_k)>\frac{1}{n}+\sum_{k=n-3}^{\frac{n}{2}}\,\frac{|G\cap\Hph_k|}{n|\Hph_k|}=
\frac{1}{n}+\sum_{k=n-3}^{\frac{n}{2}}\,\frac{\Ga(n-k)}{n\ga(n-k)}=
\frac{1}{n}+\sum_{s=3}^{\frac{n}{2}}\,\frac{\Ga(s)}{n\ga(s)}= \nonumber\\
\frac{1}{n}+ \frac{\Ga(3)}{n\ga(3)}+\frac{\Ga(4)}{n\ga(4)}+\sum_{s=5}^{\frac{n}{2}}\,\frac{\Ga(s)}{n\ga(s)}
\geq \frac{1}{n}+\frac{1}{n\al}+\Big(\frac{n}{2}-4\Big)\frac{3}{4n\al}=
\frac{1}{n}+\frac{1}{n\al}+\frac{3}{8\al}-\frac{3}{n\al}\geq 
\frac{3}{8\al},\label{eqLwBndPrG}
     \end{eqnarray}
the latter inequality follows from 
$\frac{1}{n}+\frac{1}{n\al}\geq\frac{2}{n\al}+\frac{1}{n\al}=\frac{3}{n\al}$ as $\al\geq2$. 
 
 \VSon{}Now we estimate $\Pr(G\cap\Hph_k)$, $\frac{n}{2}\leq{}k\leq(n-3)$ from above.
 From (\ref{eqBndsGa_gaSmallVal}) and lower bound on $\ga(s)$ from (\ref{eqLowBndGaga}) and 
 $\al+1\leq\frac{3}{2}\al$  we have for  $5\leq{}s\leq\frac{n}{2}$
 
   \begin{eqnarray}
   \frac{\Ga(s)}{\ga(s)} \leq \frac{1}{\al}+\frac{\al(\al+1)}{\ga(s)} \leq 
   \frac{1}{\al}+\frac{3\al^2}{2(\al^s-\al^{\frac{s}{2}+1})} = 
   \frac{1}{\al}\left(1+\frac{3\al^3}{2(\al^s-\al^{\frac{s}{2}+1})}\right) = \nonumber\\
    \frac{1}{\al}\left(1+\frac{3}{2(\al^{s-3}-\al^{\frac{s}{2}-2})}\right) =
    \frac{1}{\al}\left(1+\frac{3}{2\al^{\frac{s}{2}-2}(\al^{\frac{s}{2}-1}-1)}\right) \leq 
    \frac{1}{\al}\left(1+\frac{3}{2\sqrt{\al}(\al\sqrt{\al} -1)}\right) \leq  \nonumber\\
    \frac{1}{\al}\left(1+\frac{3}{2\sqrt{2}(2\sqrt{2} -1)}\right) \leq 
    \frac{1}{\al}\left(1+\frac{3}{2.8\cdot1.8}\right) \leq \frac{8}{5\al}\label{eqUpBndGaBy_ga}
   \end{eqnarray}
Hence for $\frac{n}{2}\leq{}k\leq(n-3)$ from (\ref{eqUpBndGaBy_ga}) and from (\ref{eqGa_ga3_4_5}) (for these formulas it is easy to check the bound directly)
   \begin{eqnarray}
\Pr(G\cap\Hph_k)= \frac{|G\cap\Hph_k|}{n|\Hph_k|}=\frac{\Ga(n-k)}{n\ga(n-k)}\leq\frac{8}{5n\al}
\label{eqUpBndPrGcap_vph}
    \end{eqnarray}

 Using the  bounds (\ref{eqUpBndPrGcap_vph}), (\ref{eqLwBndPrG}), (\ref{eqLwBndPrGcap_vph}) and $G\cap\Hph_{n-2}=\Hph_{n-2}$, we get
 
 \begin{eqnarray}
    \CD(G)= -\Pr(G\cap\Hph_{n-2})\log\frac{\Pr(G\cap\Hph_{n-2})}{\Pr(G)}-
    \sum_{k=n-3}^0\,\Pr(G\cap\Hph_k)\log\frac{\Pr(G\cap\Hph_k)}{\Pr(G)}\geq\nonumber \\
     \frac{1}{n}\Big(-\log\frac{1}{n\Pr(G)}\Big) + 
     \sum_{k=n-3}^{\frac{n}{2}}\,\Pr(G\cap\Hph_k)\Big(-\log\frac{\Pr(G\cap\Hph_k)}{\Pr(G)}\Big)\geq\nonumber\\
     \frac{1}{n}\Big(-\log\frac{8\al}{3n}\Big) + 
     \sum_{k=n-3}^{\frac{n}{2}}\,\frac{1}{n\cdot2\al}\Big(-\log\frac{8\cdot8\al}{n5\al\cdot3}\Big) =\nonumber\\
     \frac{1}{n}\Big(\log{n}-\log\frac{8\al}{3}\Big)+
     \Big(\frac{n}{2}-2\Big)\frac{1}{n\cdot2\al}\Big(\log{}n-\log\frac{64}{15}\Big)\geq
    \frac{1}{4\al}\log{}n-c(\al),\label{eqLwBndEntWghtS}
     \end{eqnarray}
 where $c(\al)$ is a `small' positive constant. 
 
  \subsection{`Small' upper bound for $H=\xi_{n-2,n-2}$}\label{UpBndXi_n-2_n-2}
 
   We have
 
   \begin{eqnarray}\label{eqCD_H_ini}
  \CD(H)= \sum_{k=0,1}\Pr(H\cap\Hph_k)\left(-\log\frac{\Pr(H\cap\Hph_k)}{\Pr(H)}\right)=
  \sum_{k=0,1}\Pr(W_k)\left(-\log\frac{\Pr(W_k)}{\Pr(H)}\right)
   \end{eqnarray}

  We look for bounds of terms of (\ref{eqCD_H_ini}) 
  
  \begin{eqnarray}
  \Pr(W_0) = \frac{|W_0|}{n|\Hph_0|} = \frac{\al(\al-1)(\al-2)}{n|\Hph_0|} <
  \frac{\al^2(\al-1)}{n|\Hph_0|},\nonumber\\
   \Pr(W_1) = \frac{|W_1|}{n|\Hph_1|} = \frac{\al(\al-1)}{n|\Hph_1|} <
   \frac{\al^2(\al-1)}{n|\Hph_1|},\nonumber\\ 
   \Pr(H)=\sum_{k=0,1}\Pr(H\cap\Hph_k) =
   \frac{1}{n}\left(\frac{|W_0|}{|\Hph_0|}+\frac{|W_1|}{|\Hph_1|}\right) < 
   \frac{\al^2(\al-1)}{n}\left(\frac{1}{|\Hph_0|}+\frac{1|}{|\Hph_1|}\right).\label{eqUpBndW} 
   \end{eqnarray}
We take very approximate bounds.  The words of the form $w=aub^{\frac{n}{2}}$, $a\neq{}b$, are in $\Hph_0$ as a prefixe of length $\leq\frac{n}{2}$ cannot be a suffixe as $a\neq{}b$, similar for bigger suffixe that gives a periodicity. 
Thus, $|\Hph_0|\geq\al^{\frac{n}{2}-1}(\al-1)$. The words of the form $w=aub^{\frac{n}{2}-1}a$, $a\neq{}b$, are in $\Hph_1$ for similar reason. So the same lower bound is valid for $|\Hph_1|$.  For $|\Hph_k|$ a trivial upper bound suffices $|\Hph_k|<\al^{n-k}$. From all these bounds, including all bounds from (\ref{eqUpBndW}) we get 

  \begin{eqnarray}
  \frac{1}{n\al^{n+1}}<\frac{\al(\al-1)(\al-2)}{n\al^n} < \Pr(W_0) <
  \frac{\al^2(\al-1)}{n\al^{\frac{n}{2}-1}(\al-1)}= \frac{1}{n\al^{\frac{n}{2}-3}}\nonumber\\
   \frac{1}{n\al^{n+1}}<\frac{\al(\al-1)}{n\al^{n-1}}<\Pr(W_1)<\frac{1}{n\al^{\frac{n}{2}-3}}
                     \label{eqBndsPW}\\
   \Pr(H)< \frac{\al^2(\al-1)}{n}\frac{2}{\al^{\frac{n}{2}-1}(\al-1)} = 
   \frac{2}{n\al^{\frac{n}{2}-3}}\label{eqUpBndPH}
   \end{eqnarray}
  From (\ref{eqCD_H_ini}), (\ref{eqBndsPW}) and (\ref{eqUpBndPH}) we get
  
  \begin{eqnarray}
  \CD(H)<-\frac{2}{n\al^{\frac{n}{2}-3}}\log\frac{n\al^{\frac{n}{2}-3}}{n\al^{n+1}\cdot2} = 
  \frac{2}{n\al^{\frac{n}{2}-3}}\log\left(2\al^{\frac{n}{2}+4}\right) = 
  \frac{2}{n\al^{\frac{n}{2}-3}}\left(\log\al^\frac{n}{2}+\log\left(2\al^4\right)\right)=\nonumber\\
  \frac{2\cdot{}n\log\al}{n\al^{\frac{n}{2}-3}\cdot2}+
   \frac{2\log(2\al^4)}{n\al^{\frac{n}{2}-3}} < \frac{\log\al}{\al^{\frac{n}{2}-3}}+
   \BO{\frac{1}{n\al^{\frac{n}{2}-3}}} = \BO{\frac{1}{\al^{\frac{n}{2}-3}}}
      \end{eqnarray}


\end{document}